# An Early In-Situ Stress Signature of the AlN-Si Pre-growth Interface for Successful Integration of Nitrides with (111) Si

Hareesh Chandrasekar,[1,a)] Nagaboopathy Mohan,[2] Abheek Bardhan,[1] K N Bhat,[1] Navakanta Bhat,[1,3] N. Ravishankar[2] and Srinivasan Raghavan[1,2]

[1] Centre for Nano Science and Engineering, Indian Institute of Science, Bangalore, 560012, India

[2] Materials Research Centre, Indian Institute of Science, Bangalore, 560012, India

[3] Department of Electrical Communication Engineering, Indian Institute of Science, Bangalore, 560012, India

The integration of MOCVD grown group III-A nitride device stacks on Si (111) substrates is critically dependent on the quality of the first AlN buffer layer grown. A Si surface that is both oxide-free and smooth is a primary requirement for nucleating such layers. A single parameter, the AlN layer growth stress, is shown to be an early (within 50 nm), clear (<0.5 GPa versus >1 GPa) and fail-safe indicator of the pre-growth surface, and the AlN quality required for successful epitaxy. Grain coalescence model for stress generation is used to correlate growth stress, the AlN-Si interface and crystal quality.

The growth of group III-A nitrides (III-Ns) – GaN, AlN and InN and their alloys – on (111) Si substrates is important due to their low cost and large size[1] as compared to the more commonly used sapphire and SiC. The growth of III-N films on Si begins with an AlN seed layer.[2] It was recognized early on that the quality of GaN layers grown on these AlN buffers are dependent upon the buffer layer deposition parameters, primarily growth temperature (900-1200°C) and thickness (10-500 nm).[2-8]

However, the presence of an oxide free and smooth Si surface favouring epitaxy also plays a major role in determining the quality of AlN buffers deposited and hence on the subsequent AlGaN-GaN epilayers. Ex-situ treatment, using HF/buffered HF, is used either as a stand-alone pre-treatment[6,9,10] or in conjunction with an in-situ thermal desorption step in a hydrogen ambient[2,3,11,12] for native oxide removal from Si. Hydrogen desorption, if overdone, also leads to etching of the Si surface.[3,13,14] Thus, as will also be shown in this paper, sub-optimal desorption results in incomplete oxide removal and excessive desorption results in surface damage. This necessitates the use of an optimum time-temperature thermal desorption step, with or without ex-situ oxide removal. In the growth of III-nitrides, this problem of optimum determination is further exacerbated by the presence of growth residues that affect reactor conditions from run to run.

A striking example of this kind of inconsistency is illustrated in Fig. 1 which shows Nomarski optical micrographs of the same high electron mobility transistor (HEMT) stacks (see Fig S1 in Ref. 15 for details of the stack) grown under identical conditions[15] after annealing the Si (111) substrate at 1050°C for 10 min in hydrogen prior to growth. The stack shown in Fig. 1(a) had a very rough surface, RMS surface roughness ($R_q$) > 100 nm, and displayed no HEMT-like behaviour. The stack shown in Fig. 1(b) had a specular surface with $R_q$ <1 nm, a 2DEG mobility of 1325 cm$^2$/Vs and a carrier concentration of $8.37 \times 10^{12}$ cm$^{-2}$. An analysis of a series of such runs indicated that the difference between the HEMT stacks lay only in the growth stress of the AlN buffer layers in both cases. Low stresses (<0.5 GPa) in the AlN buffer layer resulted in an outcome represented by Fig. 1(a) whereas high stresses (>1 GPa) yielded stacks shown in Fig. 1(b).

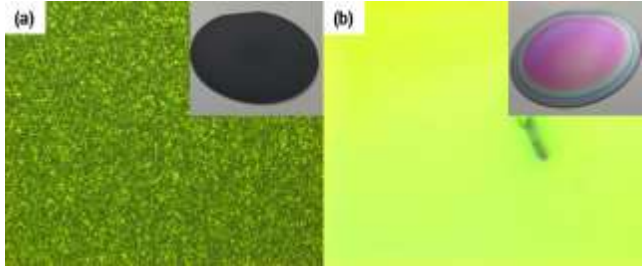

FIG. 1. 120 μm × 90 μm optical differential interference contrast micrographs of identical HEMT stacks grown on AlN buffer layers with (a) low (<0.5 GPa) stress and (b) high (>1 GPa) stress. Photographs of the corresponding 2" wafers are shown in the insets.

The correlation between surface roughness of the HEMT stack and stress in their AlN buffers is shown in Fig. 2 for 10 growth runs. Samples having >1GPa stress in the AlN buffers have an RMS roughness of around 1 nm, whereas those having low stresses (<0.5GPa) have an RMS roughness of the order of 100 nm.

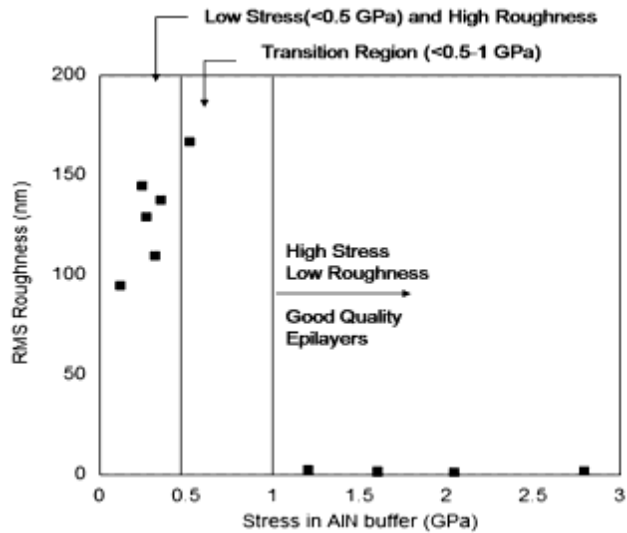

FIG. 2. Variation of RMS roughness, in nm, of the top surface of the HEMT stack with stress in the AlN buffers, in GPa, plotted for 10 growth runs. The RMS roughness is obtained from AFM scans over a 10 μm×10 μm area at the centre of a 2" wafer. Samples with stresses >1GPa have an RMS roughness of about 1 nm and those <0.5 GPa have roughnesses in the order of 100 nm.

We show in this paper that the lack of repeatability demonstrated in Figs. 1 and 2 has its origins in the inability to consistently offer an oxide free and smooth Si pre-growth surface for AlN to grow on. Given that the right surface depends not only on pre-growth treatment procedures but also on reactor drift, the availability of a growth parameter that can help assess the quality of the film in real time in every run would prove to be invaluable. The growth stress in AlN buffers is one such parameter, with a high growth stress, >1 GPa, being an early and reliable representative, as evident from Fig. 2, of both the optimally treated interface and the AlN crystalline quality required for successful integration with Si.

In order to understand the correlation between AlN stress and the pre-growth surface, the temperature-time combinations for in-situ thermal annealing of Si substrates in hydrogen was studied.[15] Atomic force microscopy (AFM) of the (111) Si surface following such exposure revealed significant surface pitting for samples annealed at 1050°C-10 min and 1050°C-5 min (Figs. 3(a) and 3(b), also see Fig. S2 in Ref. 15 for SEM images). It is to be noted that 1000-1200°C is a typical AlN growth window. This observation is in agreement with earlier reports on surface roughening of Si before III-N growth,[3] though the reason for this pitting remains unclear and beyond the scope of this paper. A reduction in the time of desorption to 0 minutes at 1050°C (Fig. 3(c)) or the temperature to 900°C (Figs. 3(d)-(f)) eliminated any pitting and resulted in smooth surfaces that had no detectible difference in surface roughness from as received Si wafers.

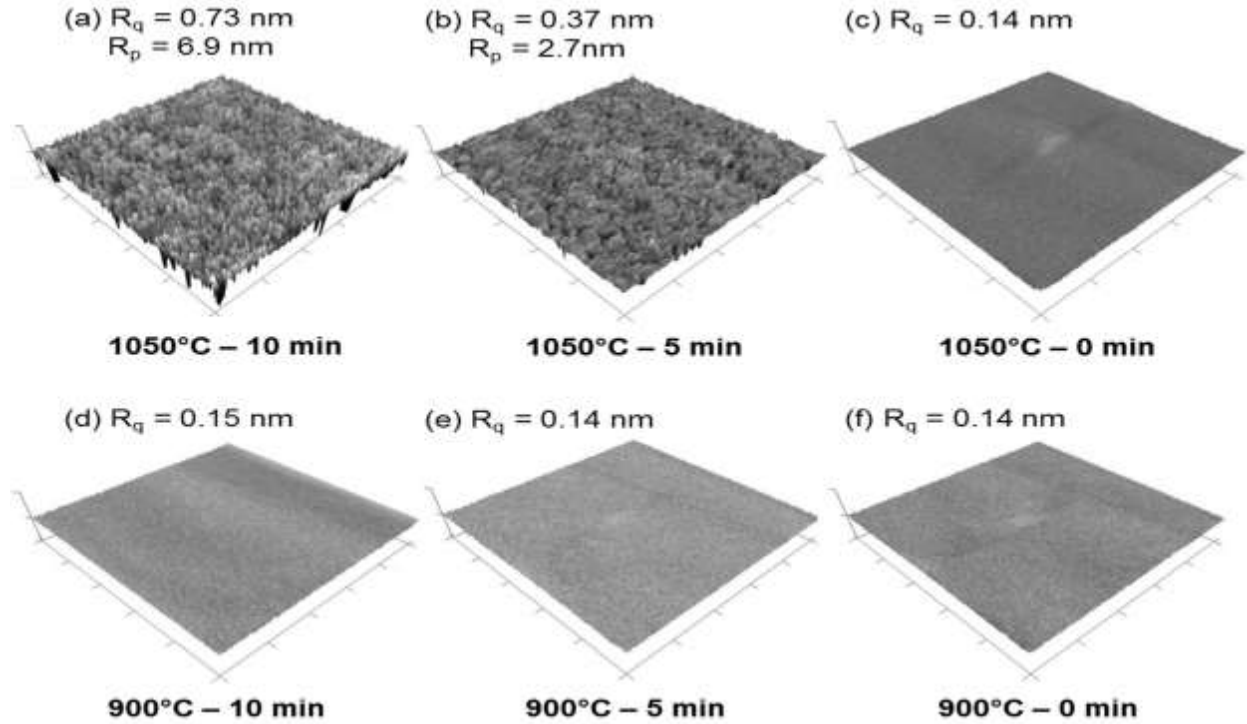

FIG. 3.(a)-(f) 10 μm x 10 μm AFM scans of the Si (111) surface after thermal desorption using temperature-time combinations listed alongside the figures. The z-scale in all cases is 5 nm. $R_q$ is the RMS roughness. For the 1050°C-5 min and 1050°C-10 min cases, $R_p$ represents the average pit-depth.

Since thermal desorption is done primarily to remove the native oxide layer, it is obvious that the conditions giving smooth Si surfaces need to be evaluated for their efficacy of oxide removal as well. Both stand-alone in-situ desorption and, ex-situ oxide removal (using a high purity buffered HF solution) in conjunction with in-situ desorption were investigated. Both the as-received and ex-situ treated wafers were immediately loaded into the growth chamber and thermally desorbed under conditions that did not cause surface pitting; 1050°C-0 min, 900°C-0 min and 900°C-10 min; prior to AlN deposition at 900°C.

Fig. 4 shows the stress in the AlN buffers measured for multiple growth runs after various pre-treatment conditions. The stress in the AlN layers was extracted from the slope of the stress-thickness vs thickness plots (see Fig. S3 in Ref. 15) obtained from in-situ measurements. It can be seen that even though all the data in Fig. 4 represented by circles, is from AlN grown at 900°C, there is a huge variation in stress from as small as 0.2 GPa to as large as 3 GPa depending on the pre-growth treatment the Si substrate was subjected to. The key finding of this paper is the recognition that the nature of this pre-growth Si surface is very important in determining the crystalline quality and hence stress in the AlN. Given the correlation established in Fig. 1 and 2 between AlN layer stress and the success of subsequent growth the importance of this observation is obvious. For the purposes of discussion, we broadly classify these into low stress (<0.5 GPa) and high stress (>1 GPa) AlN.

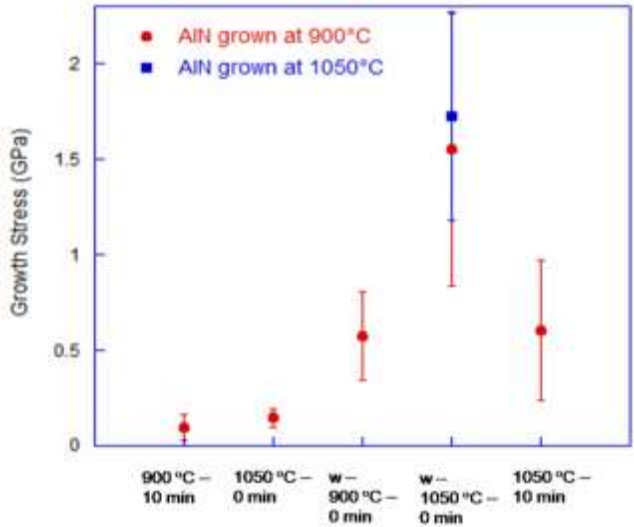

FIG.4. Mean stress in the AlN layers grown after different Si (111) pre-growth treatments indicated along the x-axis. The prefix w refers to samples that were wet etched in addition to the thermal desorption indicated. The error bars are indicative of range and are plotted over 3 runs for all the conditions, except for AlN grown after the 1050°C-10 min anneal where the standard deviations are plotted for 10 runs, and over 15 runs for the 1050°C grown AlN.

The following points which can be inferred from Fig. 4 are noteworthy.

1. The low stress in the AlN layers even after desorption at both 900°C-10 min and 1050°C-0 min indicates - the need for a minimum desorption temperature, > 900°C, and the inadequacy of desorption alone to obtain AlN of high quality.

2. While wet etching the sample prior to growth results in a jump in the measured growth stress to 0.6 GPa) in samples exposed to 900°C-0 min prior to growth, this is still much lesser than the 1 GPa regime indicative of good quality growth.

3. The steep increase in growth stresses to >1 GPa for AlN layers grown under a combination of in-situ desorption at 1050°C-0 min and ex-situ oxide removal is indicative of a substantial improvement in AlN texture and represents the optimal treatment to obtain good quality layers.

4. AlN layers grown at 1050°C after optimal pre-treatment show higher stress and a lower standard deviation as compared to the 900°C buffers. This is in keeping with earlier reports on the variation of AlN stresses with growth temperatures[16] and shows that AlN buffer layers can also be grown under this temperature reliably.

5. The stress levels in AlN layers grown after desorption at 1050°C-10 min, which causes surface pitting, span both the high and low stress regimes indicative of the run-to-run variability alluded to earlier.

In summary, a combination of wet chemical etching and optimum thermal exposure is required to obtain AlN with the highest tensile growth stresses. We now show that this optimum combination essentially results in an oxide free and smooth surface.

In order to understand the jump in AlN growth stress on wet etching, cross-sectional high resolution transmission electron microscopy studies were performed on two samples, with and without wet etching. In both cases the samples were exposed to 1050°C-0 min prior to growth at 900°C. Growth stresses in the two samples with and without wet etching were 0.18 and 2.4 GPa respectively.

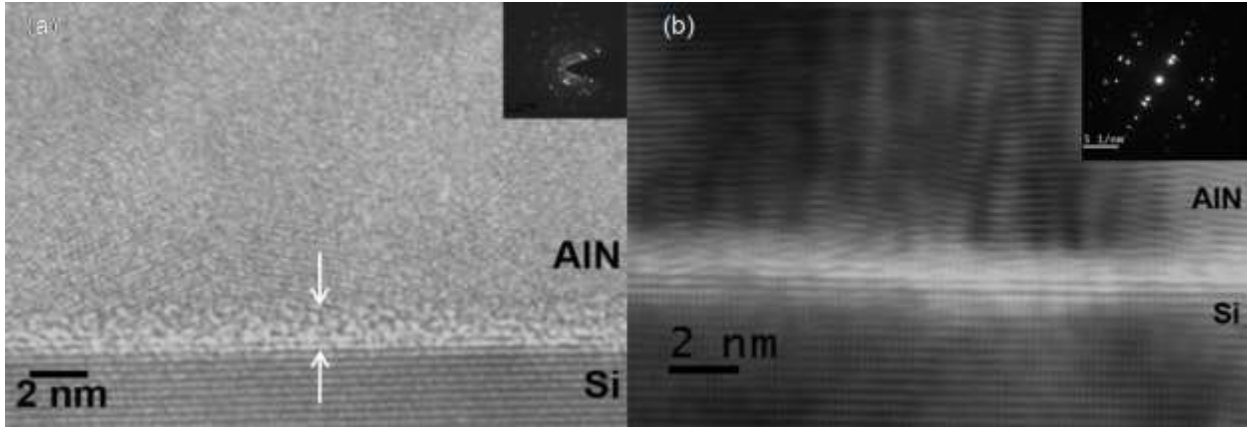

FIG.5. Cross sectional HRTEM image of (a) the non wet etched low stress AlN-Si interface and (b) wet etched high stress AlN-Si interface. HR imaging is done at the $(10\bar{1}0)$ zone axis. An amorphous region of about 2 nm is visible at the Si-AlN interface for the low stress case

In the case of the low stress sample in Fig 5(a), an amorphous region, indicated by the arrows, is observed over the entire electron transparent lamella and extends to a few monolayers of thickness. In clear contrast, in the high stress sample, Fig. 5(b), the interface between Si and AlN is a sharp one and practically devoid of any amorphous layer. The diffraction pattern from the high stress sample revealed a single crystalline spot pattern shown in inset of Fig 5(b), whereas the low stress AlN sample exhibits a discontinuous ring that is characteristic of polycrystalline nature (inset of Fig 5(a)). This result in conjunction with the previous discussion clearly establishes the correlation between incomplete oxide removal, poorly oriented AlN and low growth stress.

To demonstrate the effect of sub-optimal pre-treatment on crystal quality (incomplete removal of oxide), a probe layer of 500 nm of GaN was grown directly on top of both high and low stress AlN buffers. HR-XRD rocking curves for the (0002) GaN reflections have FWHMs of 0.338° and 4.91° for the low and high stress buffers respectively, a trend that was also observed for the $(10\bar{1}1)$ reflections (see Fig. S6 in Ref. 15). A similar effect is also observed for low and high stress AlN obtained by super-optimal desorption with the GaN reflections from the HEMT stack, shown in Fig. 1, being used for HR-XRD analysis, with layers on low stress buffers displaying much higher FWHMs compared to those on high stress AlN (see Fig. S7 and S8 in Ref. 15).

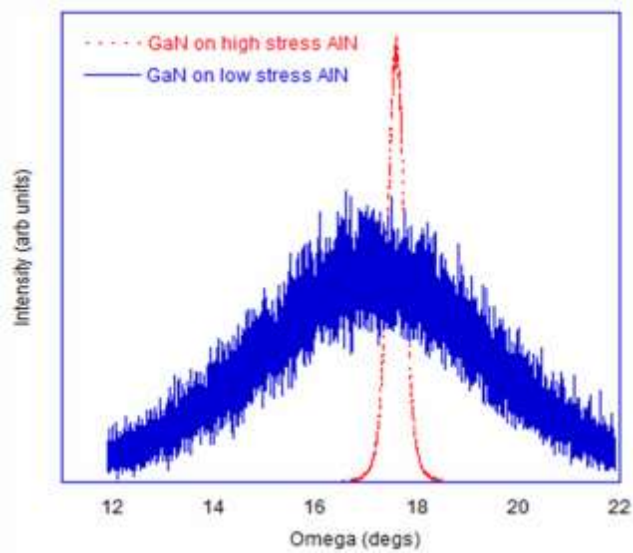

FIG. 6. HR-XRD (0002) rocking curves from GaN films grown on high stress and low stress AlN buffers obtained under sub-optimal oxide removal conditions.

AlN films grown on Si are expected to have a compressive stress of about -5 GPa based on the epitaxial 4-5 arrangement ($4\times a_{Si} \approx 5\times a_{AlN}$) with misfit segments being generated. However, as has been documented elsewhere,[16-20] only a constant tensile stress is observed right from the beginning of AlN growth on Si. This is particularly true of low mobility films grown at relatively high rates in excess of 1 nm/sec as is typical of MOCVD. It has been shown that at such high growth rates the maximum tensile stress, especially at the uppermost layers, becomes independent of the growth rate and can reach the theoretical cohesive strength of the boundary, which scales as $(2\gamma_{sv} - \gamma_{gb})^n$.[18] $\gamma_{sv}$ and $\gamma_{gb}$ are the solid-vapour interfacial energy and the grain boundary interfacial energy respectively. "n" is a positive exponent and typically equal to 0.5. Hence, the lower stress with poorer texture, such as in the polycrystalline AlN films, is due to the higher $\gamma_{gb}$ of the boundary formed and vice versa.[17]

We wish to emphasize that while both low (<0.5 GPa) and high (>1 GPa) stress regimes have previously been observed due to growth temperature variations during AlN growth,[16,17] the fact that AlN films in all our cases are grown at the same temperature of 900°C allows us to isolate the effect of the substrate on crystalline quality and therefore stress generation in these films.

In conclusion, the importance of the AlN-Si interface in obtaining reproducible good quality III-N epitaxial layers on Si substrates has been discussed. A smooth, oxide free Si surface is shown to be crucial for the growth of good quality AlN buffers. An optimum time-temperature combination for in-situ thermal desorption, allied with ex-situ oxide removal is shown to be essential to obtain such interfaces. In-situ stress measurement during growth of the AlN buffers is shown to provide a fail-safe, effective and timely signature (within 50 nm of AlN growth) of the presence of such an interface, the quality of the AlN film itself and the epilayers grown on it. Low tensile stresses (<0.5 GPa) correspond to poor material quality and high stresses (> 1 GPa) correspond to good quality layers.

We acknowledge the Ministry of Defence, GOI through sanction number TD-2008/SPL-147 for funding to carry out this research, the Micro and Nano Characterization Facility of the Centre for Nano Science and Engineering, Indian Institute of Science for providing access to AFM, SEM, and XRD and the Advanced Facility for Microscopy and Microanalysis of the Indian Institute of Science for access to the TEM. We would also like to acknowledge fruitful discussions with Dr. Assadullah Alam, AIXTRON AG, Germany. H.C. wishes to thank Mr. Y G Radhakrishna for assistance with the AFM imaging.

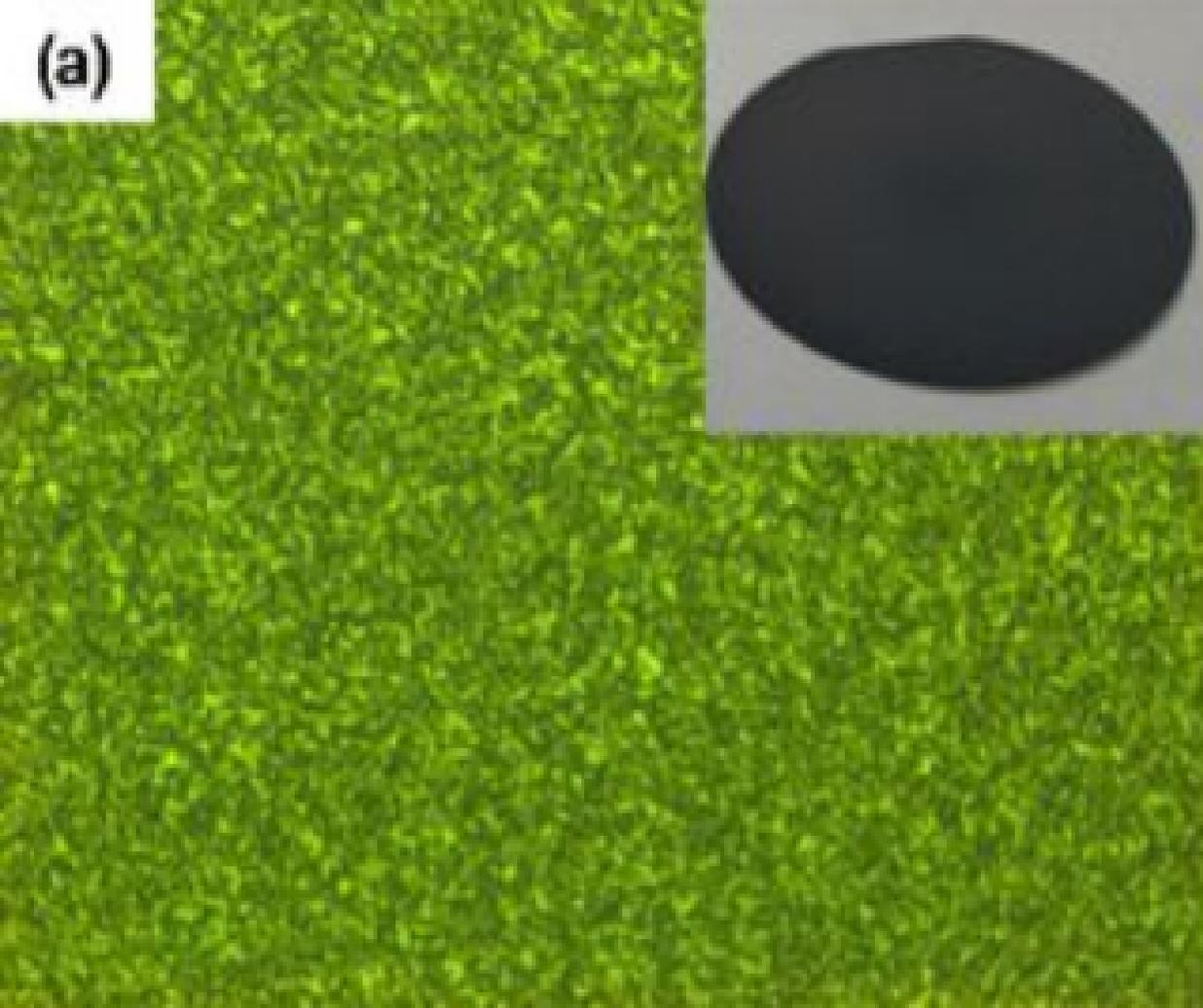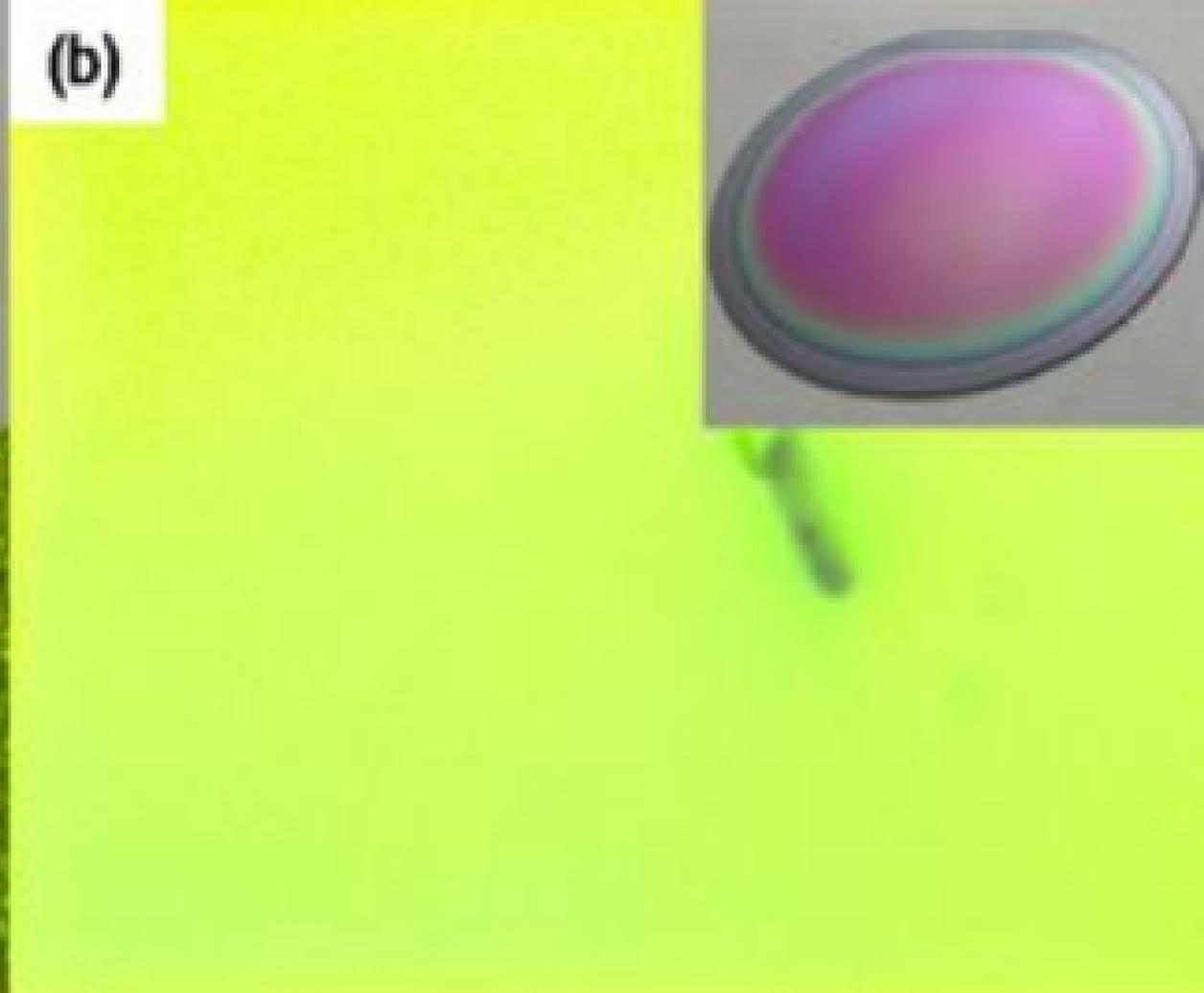

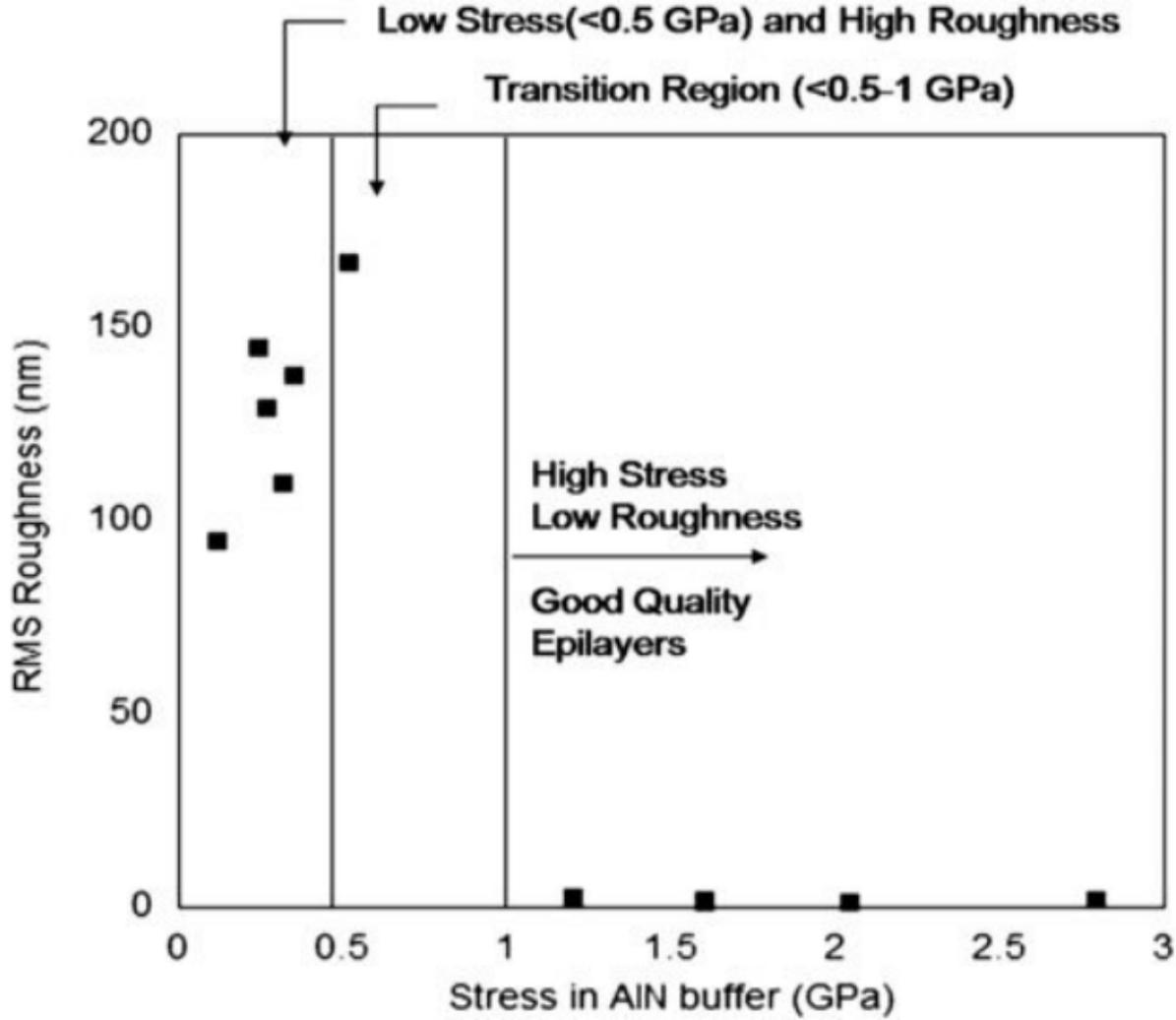

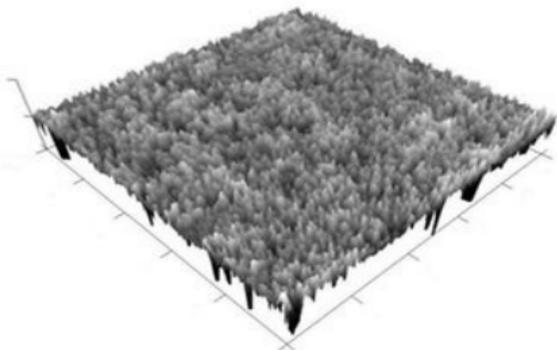 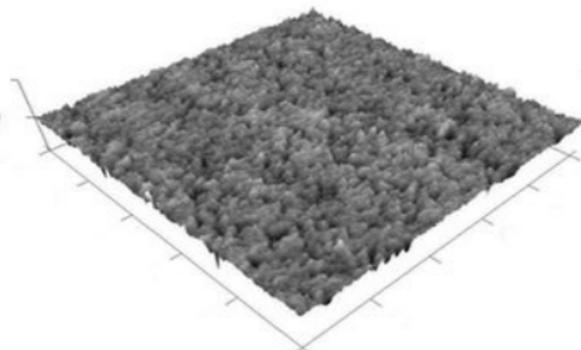 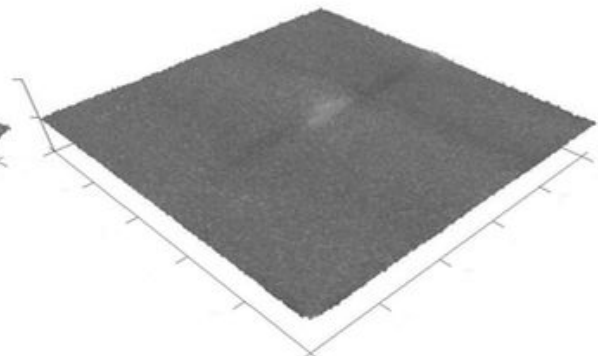
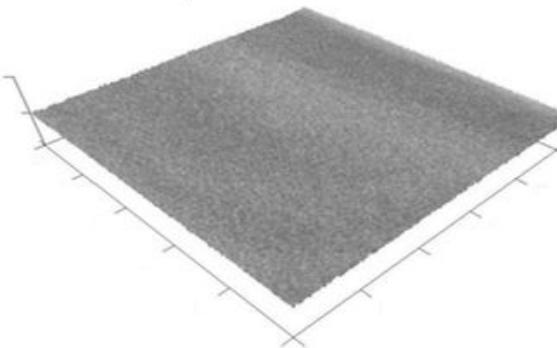 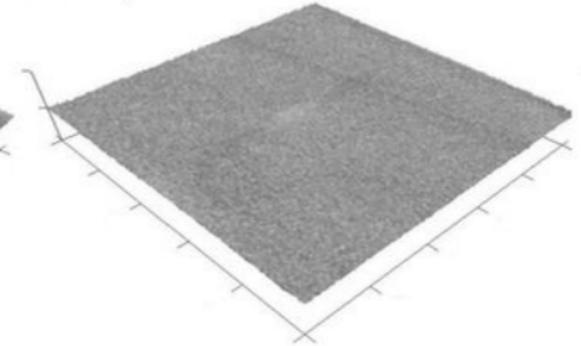 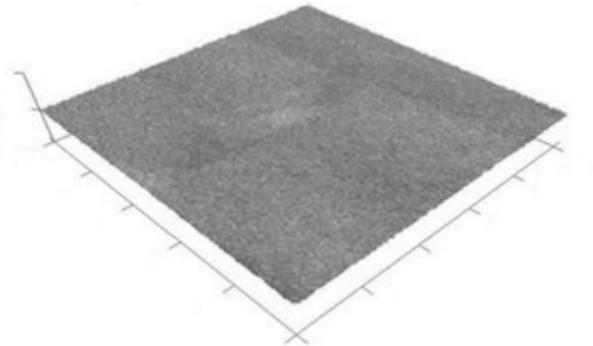

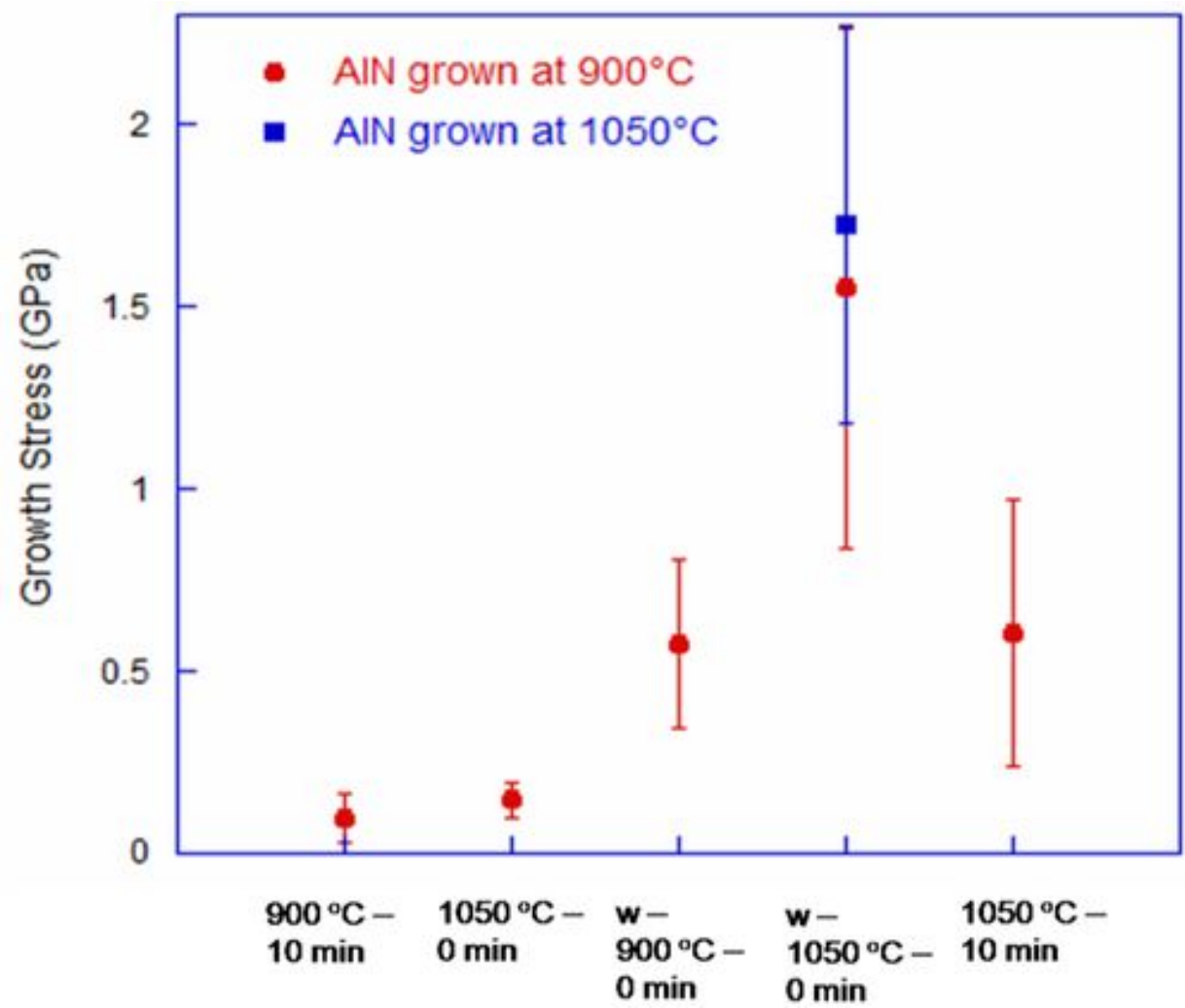

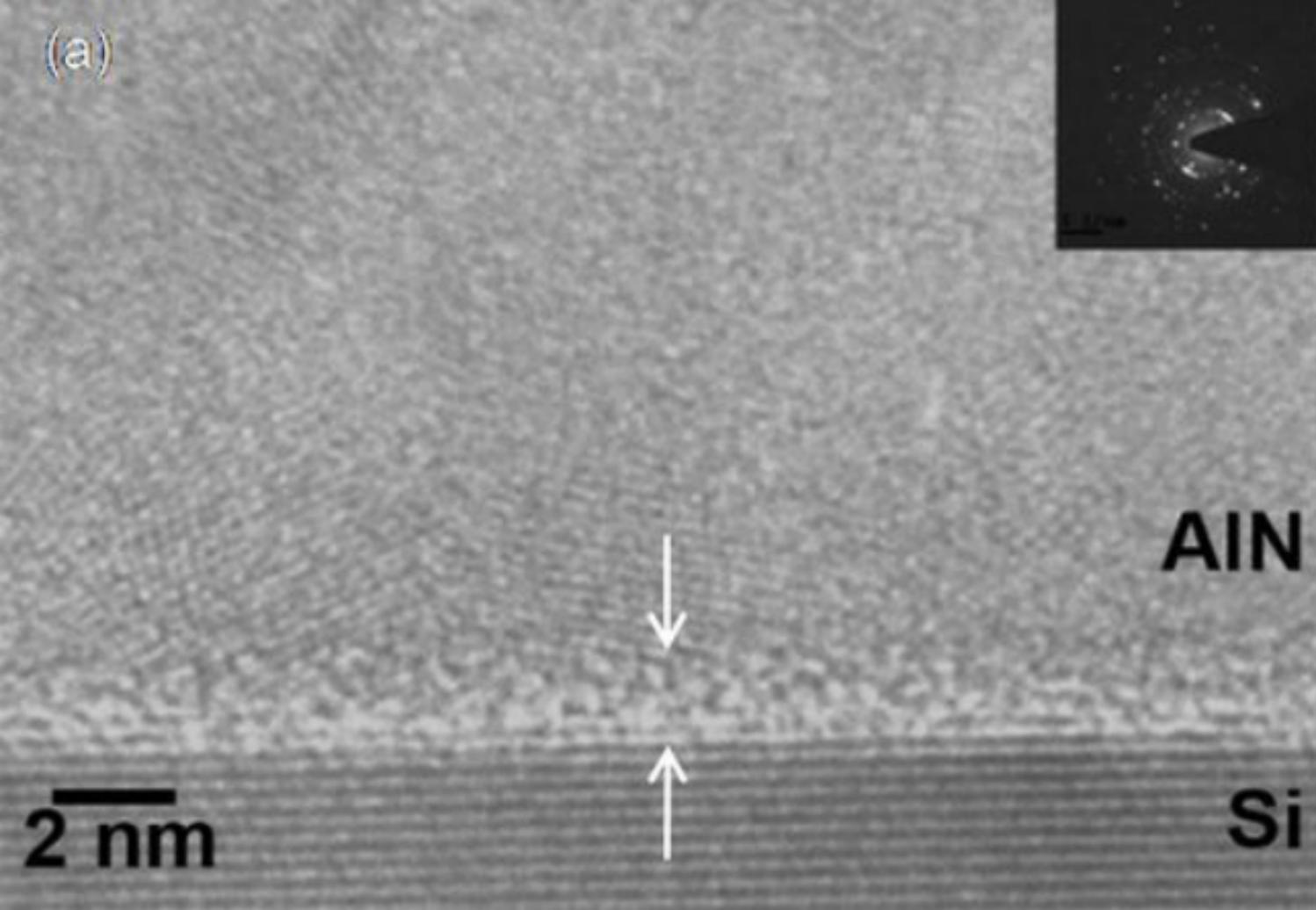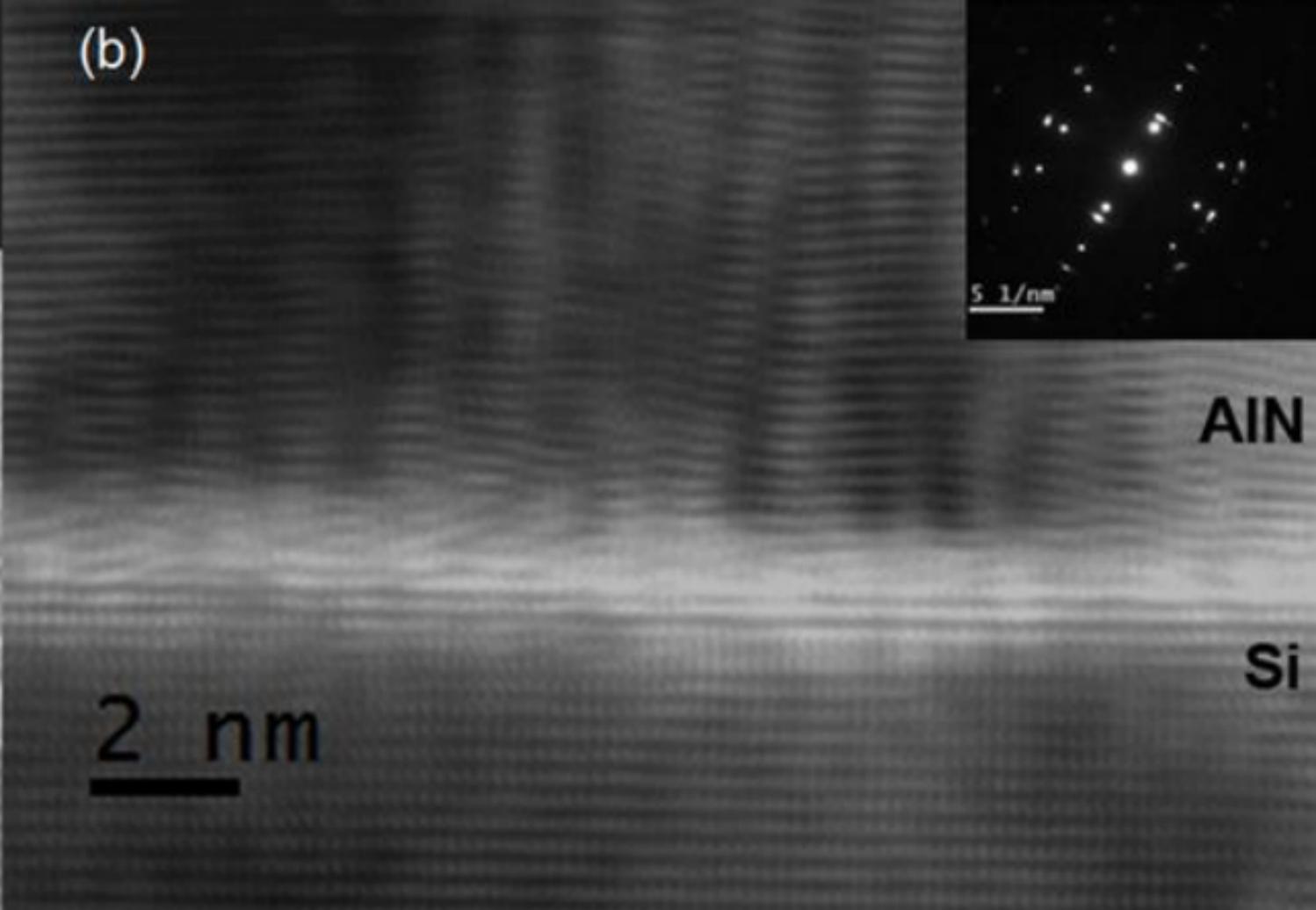

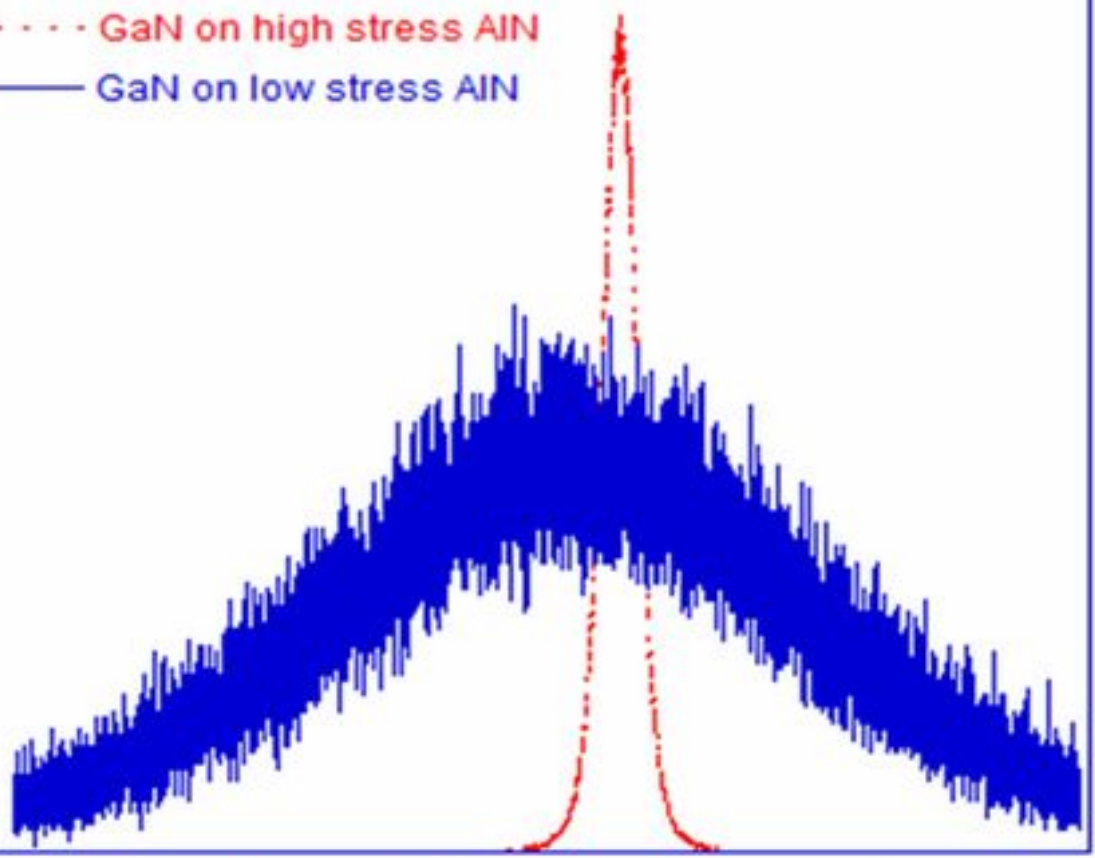